# Towards an AI/ML-defined Radio for Wi-Fi: Overview, Challenges, and Roadmap

Boris Bellalta[1], Katarzyna Kosek-Szott[2], Szymon Szott[2], Francesc Wilhelmi[3]

(1) Universitat Pompeu Fabra, Barcelona, Spain
(2) AGH University of Krakow, Poland
(3) Nokia Bell Labs, Stuttgart, Germany

**Abstract**

Will AI/ML-defined radios become a reality in the near future? In this paper, we introduce the concept of an AI/ML-defined radio—a radio architecture specifically designed to support AI/ML-based optimization and decision-making in communication functions—and depict its promised benefits and potential challenges. Additionally, we discuss a potential roadmap for the development and adoption of AI/ML-defined radios, and highlight the enablers for addressing their associated challenges. While we offer a general overview of the AI/ML-defined radio concept, our focus throughout the paper remains on Wi-Fi, a wireless technology that may significantly benefit from the integration of AI/ML-defined radios, owing to its inherent decentralized management and operation within unlicensed frequency bands.

## 1. Introduction

Artificial intelligence (AI) can be seen as the machine-like implementation of the 'human cognitive cycle': gathering data, extracting knowledge from the gathered data, reasoning and actuating based on that knowledge, and finally building a casual relationship between the actions taken and the results obtained. In such a 'cognitive cycle', machine learning (ML) models are in charge of extracting and exploiting knowledge from the data gathered, which is retained in a 'learned from experience' function. For example, neural networks (NNs) are commonly used for image classification, which is achieved by finding specific patterns in the data [1].

In wireless communications, ML models should likewise enable radio interfaces to extract knowledge from gathered data and guide future decisions. AI/ML has already been identified as a key enabler in 5G and 5G-Advanced cellular networks for network management and orchestration, thus leading to enhanced ways of planning networks, providing troubleshooting, or optimizing performance [2]. For Wi-Fi, the literature review in [3] shows how AI/ML can be applied for improving channel access, link configuration, and beamforming, among many other Wi-Fi features. In addition, a deep dive into the evolution of the IEEE 802.11 protocol towards native AI operation is given in [4] while for a more general discussion in the context of 6G, we refer readers to the Hexa-X project vision on AI/ML-native radios [5].

Following the promising results observed when AI/ML is used to improve wireless communication, it naturally raises the question of how a radio designed with support for AI/ML can be built, and if such a design would outperform expert-based designs. Nokia's pioneering vision on an AI/ML-native air interface, presented in [6], is a comprehensive reference to this topic and lays down the fundamentals for next-generation communications systems to embrace these concepts. This reference overviews the benefits of employing AI to AI-radio ends, including several examples of how physical layer features can be replaced by ML models, such as learning the symbol constellation to allow pilotless transmissions when a neural receiver is employed. Today, the AI/ML-native air interface is becoming a reality thanks to the proof-of-concepts and implementations done (including interoperability tests) towards the 6G air interface, which is foreseen to include AI-generated waveforms and AI-based receivers able to process them.[1,2]

In this paper, we provide insights into the potential functions of a future AI/ML-defined radio interface in Wi-Fi and devise potential steps for adoption. Wi-Fi poses unique challenges when it comes to adopting AI/ML-defined radio principles—at least for link-layer functionalities—due to its decentralized nature and operation in unlicensed bands. Accordingly, the interaction of a Wi-Fi device with the environment significantly differs from that of cellular communications, provided that neighboring interference and access to the channel are not controlled in Wi-Fi. In such a context, it could be relevant that each network and/or device supports autonomous and self-configuration capabilities based on AI/ML to adapt the communication process to different and changing environments where, on top, a plethora of heterogeneous capabilities can be found. Following the works surveyed in [3], significant performance gains are achieved thanks to implementing ML models in different Wi-Fi functionalities. However, those gains are obtained in most of the cases under very specific and ideal assumptions, without considering practical and implementation aspects, which in our opinion represent the major bottleneck in the roadmap towards AI/ML-defined radios.

Now the question is, against this background, when can we expect to see AI/ML-defined radios in Wi-Fi? While Wi-Fi 8, developed from the IEEE 802.11bn amendment [7] and expected for 2028, will likely be complemented with some proprietary AI/ML features developed at the vendor side (e.g., to smartly handle multi-access point coordination), it is not expected that the IEEE 802.11 standard will include any AI/ML-related functionalities before Wi-Fi 9. Meanwhile, as we await the outcome of such standardization efforts, we use a clean-slate approach to define a future AI/ML-defined radio interface (Section 2), outline a roadmap towards AI/ML-defined radios (Section 3), and summarize our views in Section 4.

---

[1] https://www.nokia.com/about-us/news/releases/2024/02/22/nokia-skt-ntt-and-docomo-team-up-to-implement-ai-in-the-6g-air-interface/
[2] https://www.nokia.com/about-us/news/releases/2024/02/20/nokia-and-qualcomm-jointly-research-ai-interoperability-technology-that-boosts-wireless-capacity-and-performance/

## 2. AI/ML-defined Radio

In this section, we provide a definition of AI/ML-based radios, list the supported functionalities, and address open challenges.

### 2.1. What is a Wi-Fi AI/ML-defined Radio Interface?

An AI/ML-defined radio natively supports AI/ML-based optimization and decision-making in its communication functions (such as data transmission and reception, channel access and frame exchange, and radio resource management). With such features, an AI/ML-defined radio is expected to improve the communications efficiency, reduce the existing overheads, and make better use of the spectrum resources. Also, such an interface will enable more efficient transmitter and receiver implementations and higher flexibility in diverse and changing scenarios.

From an architectural standpoint, AI/ML-defined radios diverge from traditional systems that are optimized for specific, predefined firmwares. AI/ML-defined radios should be designed around or at least incorporate ML accelerators to enable rapid ML model inference. Additionally, if on-device training is required, additional storage and computational resources are needed to avoid compromising transmit/receive (TX/RX) functionalities. Finally, regarding the design of AI/ML-enabled radios, although fully flexible architectures theoretically offer the advantage of supporting continuously evolving AI/ML models, it is more probable that we will see circuits optimized to efficiently execute only specific types of ML models. This specialization, however, may limit the radio's ability to quickly adapt to emerging ML models. Figure 1 illustrates the modular concept of an AI-defined radio, wherein ML agents and models govern the radio processing resources and radio frequency (RF) hardware, the frame exchange between the transmitter and the receiver, and the signals exchanged over the air.

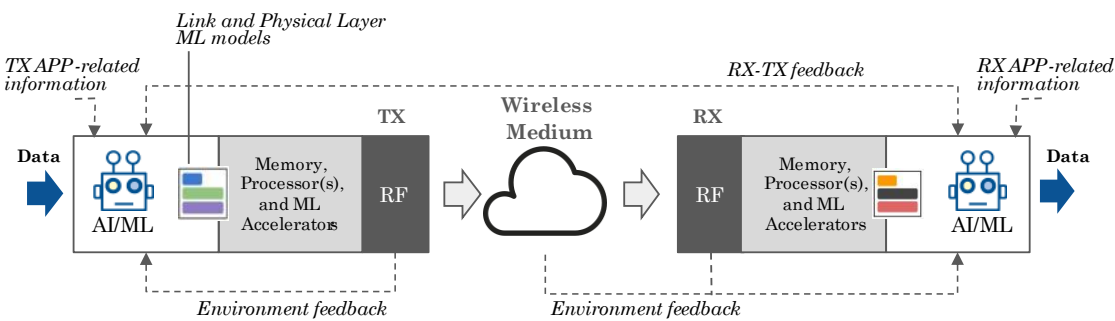

Figure 1. The AI-defined radio concept.

## 2.3. Wi-Fi AI/ML-defined Radio Functionalities

Wi-Fi radio functionalities can be organized into two groups: physical layer (data TX/RX pipeline), and link layer (interactive protocols and radio resource management). In the following, we provide some examples of radio functionalities and explain how they can resort to AI/ML to improve—or even replace—their operation.

**1) TX/RX pipeline (physical layer)**: The physical layer functionalities operate on the transmitted and received wireless signals, also referred to as the air interface. At the transmitter side, functionalities like channel coding, modulation constellation design, and mitigation of RF hardware impairments can be enhanced or replaced by ML models. Similarly, at the receiver side, we encounter functionalities such as channel estimation, signal equalization, symbol demapping, and decoding. While replacing individual functions can lead to significant performance improvements, such as achieving comparable performance at lower signal-to-noise ratio (SNR) values, even greater benefits can be realized by employing ML models to replace multiple functions and facilitate joint optimization of transmitter and receiver functionalities [6]. For a comprehensive introduction to how ML models can replace physical layer functionalities, we recommend [8, 9] where examples to jointly optimize the transmitter and receiver are provided.

**2) Interactive protocols and radio resource management (link layer)**: Link layer functionalities manage channel access and facilitate the exchange of data and metadata, such as channel and buffer state feedback, between the transmitter and receiver using communication protocols (Figure 2). Link layer operations entail high complexity and AI/ML can play an important role in addressing the arising challenges [10]. Moreover, in the particular case of Wi-Fi, link layer actions taken by one device significantly influence the actions of others, largely due to the mandatory listen-before-talk (LBT) spectrum sharing operation. This scenario makes employing ML agents to learn optimal strategies for interacting with the environment an interesting option [11]. Representative link-layer functionalities in Wi-Fi include:

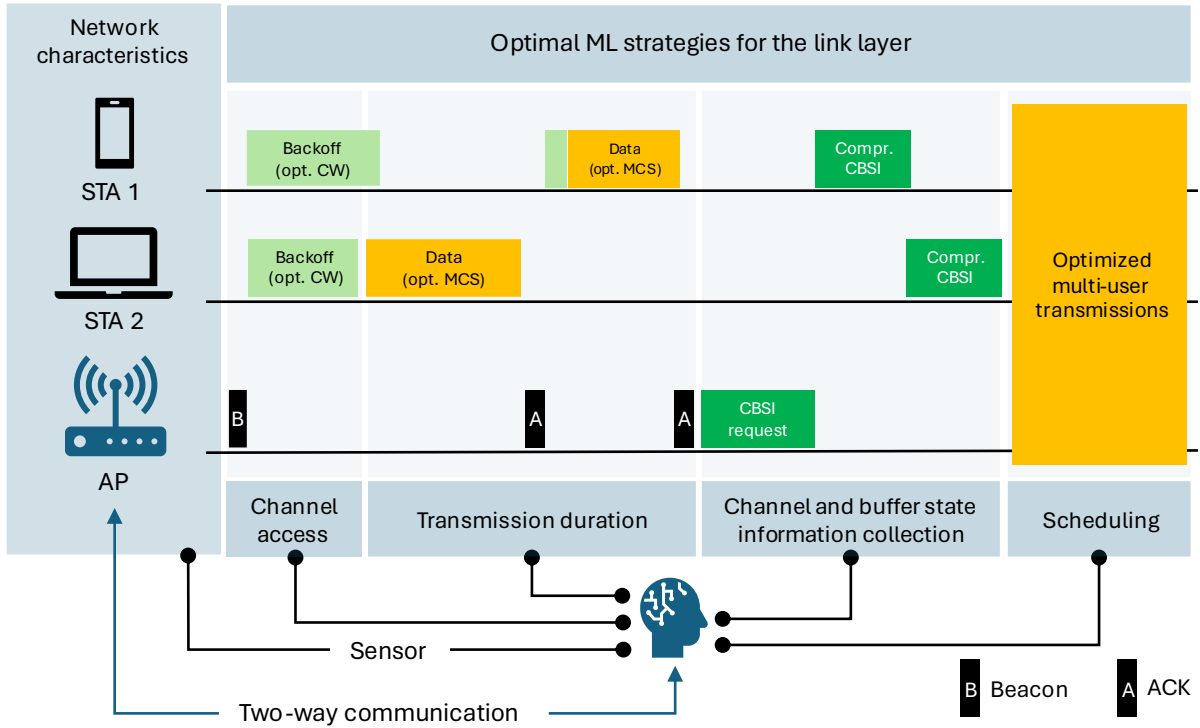

Figure 2. ML-supported Wi-Fi link layer.

- Channel access: ML models can be used to predict when a device should defer or contend for the channel in the presence of potential interferers (i.e., spatial reuse opportunities), as well as to set the most suitable channel contention parameters (e.g., CW values) for any particular scenario. A particular example is the use of a deep Q-network (DQN) for CW optimization [12]. In a more futuristic stage, AI/ML can be employed to directly drive the channel access, thus relying on the ML model to make the necessary decisions.

- Transmission duration set-up: After determining channel access, the transmission duration must be decided based on the volume of data waiting for transmission and a feasible modulation and coding scheme (MCS) available. ML models can assist in these decisions by addressing the tradeoff between maximizing the efficiency of large transmissions and minimizing the increased delays imposed on other contending devices that must defer. A particular example showing how NNs are able to select the proper MCSs by identifying the correlations between MCS, link quality, and achievable performance is presented in [13].

- Channel and buffer state collection: Based on the application requirements and scenario characteristics, ML agents deployed at the AP can learn which is the optimal rate to request channel and buffer state information (CBSI) from their associated stations and other APs. Additionally, ML models, such as autoencoders, can compress data to reduce channel training and feedback exchange overheads. This use case is addressed by the 802.11 AIML Topic Interest Group (TIG) [14].

- Scheduling: Wi-Fi technology provides numerous degrees of freedom for device and traffic scheduling, such as quality of service (QoS), orthogonal frequency division multiple access (OFDMA), multi-user multiple-input and multiple-output (MU-MIMO), target wake time (TWT), multi-link operation (MLO), and multi-access point coordination (MAPC). However, this versatility comes with increased complexity in managing these features comprehensively. AI/ML solutions offer promising directions to address this complexity effectively. For example, a first key challenge to leverage coordinated spatial reuse (C-SR) and coordinated beamforming (C-BF) mechanisms—currently under discussion for IEEE 802.11bn—is clustering devices and users based on their traffic requirements, observed environment and technological capabilities, specifically to handle their ability to transmit and receive data simultaneously. In [15], although not considering Wi-Fi, the authors show how unsupervised learning techniques can be leveraged to create groups of users that can be scheduled together for MU-MIMO transmissions.

While not directly related to wireless data communication, other features can benefit from the adoption of ML models. In channel selection, ML agents implementing reinforcement learning (RL) algorithms can gather and utilize information about available channels [16], thus leading to policies that are beneficial in the long term. Moreover, ML models can further exploit detailed channel state information to identify sources of interference, including characteristics of contending signals based on received data [17]. Similarly, for AP selection and roaming, stations can gather information about nearby APs, such as load and received signal level. When combined with mobility patterns, this data can suggest optimal handover times [18].

Finally, future Wi-Fi networks will likely integrate sensing capabilities providing information about the surrounding environment, as planned in the IEEE 802.11bf amendment [19]. Such information will be leveraged by new applications (motion and presence detection, fall monitoring, etc.) but can also be used to improve communication performance by incorporating a rich 'physical' context. ML models can be naturally used for sensing, i.e., to extract patterns and knowledge from observations.

All the aforementioned features have interrelated aspects that should be considered collectively, as decisions made by any individual functionality can impact others. However, constructing a single, large ML model capable of simultaneously considering all available degrees of freedom is currently impractical due to its complexity. To address this challenge, in addition to decomposing the problem into manageable subproblems, AI/ML-defined radios will necessitate an ML-aware coordination framework able to select and parameterize the required set of communication functionalities for any given traffic flow/application[3], as well as for sharing information and decisions among the various ML-enhanced functionalities. This is one of the goals of the ongoing CHIST-ERA MLDR ("An ML-Defined Radio interface") project.[4]

## 2.3. Open Challenges

Before AI/ML-defined radios become a reality, performance gains must be consistently observed. For example, using AI/ML-generated bespoke waveforms can lead to lower error rates. Similarly, by improving the exchange of channel state information (CSI) and other metadata, overheads are to be reduced.

However, AI/ML-driven radios should also be evaluated considering their adaptability to different situations, which means they will have to provide the required service levels in many more situations than nowadays, with few to no exceptions. Therefore, to evaluate an AI/ML-driven radio, in addition to raw performance, we will have to consider aspects such as training time and generalization capabilities to different situations. Would we tolerate a lower peak performance in some particular cases but higher adaptation capabilities resulting in a good but not excellent performance everywhere?

In addition, AI/ML-defined radios will need to perform regular model exchanging, deployment, and retraining to update the ML models in place. This, in turn, requires additional infrastructure and processes to be held both in the radio itself (e.g., low-cost model inference) and in cloud/edge equipment (e.g., data storage, model training, distributed learning orchestration). Moreover, considering that data is the fuel of ML application, a huge amount of ML data, including not only training data but also ML model data and metadata, will be required to be exchanged throughout different parts of the network. In this regard, assuming that part of this data will be exchanged through Wi-Fi, new Wi-Fi access categories will likely be extended to include the special case of ML-related traffic.

Likewise important is the trust placed in ML models, which is challenging as they often act as black boxes. An appealing solution to enforce trust in AI/ML lies in sand-boxes and digital twins, which allow training and evaluating ML models in a safe environment. ML sand-boxes and digital twins are likely to become widespread due to the large resources available in the cloud [20]. Those tools, however, require to accurately replicate the real environment and scenario, as well as to exchange large amounts of data between the network and the cloud in real-time. As a result, they pose an important challenge when it comes to representation accuracy versus timeliness.

---

[3] Should the same functionalities—or the same parameters—be used to support interactive Virtual Reality applications than those for fast file exchange? Likely not.
[4] https://www.upf.edu/web/mldr.eu

Finally, another significant challenge lies in achieving interoperability between devices that implement different ML models, which may necessitate deploying a standard ML model by default in all functionalities, as it could be the case for CSI exchange.

## 3. Roadmap Towards AI/ML-defined Radios

When will AI/ML-defined radios be a reality? Although it is hard to make a precise prediction, we foresee they will become the standard during the 2030s. To reach that moment, joint partnerships between the scientific and engineering, as well as the industrial and standardization communities are required.[5]

### 3.1. AI/ML-defined Radio Evolution

ML models are already deployed to support management functions. For example network traffic and user mobility prediction can support turning APs on/off, thus reducing the energy consumed by the network [21]. However, all such functionalities are either executed in the cloud, or in a capable network controller. Focusing only on ML operations executed at the radio interface, we expect the following phases (Figure 3):

1. **(Phase 1) ML models enhancing current functionalities**: Low-complexity ML models support link-layer functionalities such as channel access, transmission power adjustment for spatial reuse, user grouping and scheduling, and CSI compression.

2. **(Phase 2) ML models replacing traditional radio functionalities**: With the availability of chips supporting ML operations, we expect to see ML models replacing traditional radio functionalities at the physical layer. Neural transmitter and receiver designs able to overcome hardware impairments, working at lower signal to noise ratios (SNRs) and supporting data exchange with lower overheads (i.e., pilotless), would further improve communications efficiency. Moreover, link-layer functionalities can also be further enhanced with more complex and accurate ML models.

3. **(Phase 3) AI/ML-defined radio interface**: Radio interfaces that go beyond employing ML models, able to build and coordinate their own communication stack by deciding not only which functionalities are required but also how their state-machine is implemented to optimally serve a certain application. We expect a large interplay between semantic communication, AI/ML-driven transport and network protocols and AI/ML-defined radios, enabling a new communication era.

---

[5] A first step towards AI/ML-defined radios can be seen in the recent announcement by Qualcomm of its newest Wi-Fi 7 chip with integrated AI/ML functions for application/traffic classification: https://spectrum.ieee.org/wifi-7-qualcomm

As noted in [3] and elaborated upon in the preceding section, numerous studies demonstrate that ML models often yield improved performance. Nonetheless, there is a notable absence of publications presenting results from AI/ML-defined radio prototypes, even in simplified forms. Developing ML models optimized for execution on specialized ML chips represents a promising research and innovation avenue expected to advance significantly in the coming years.

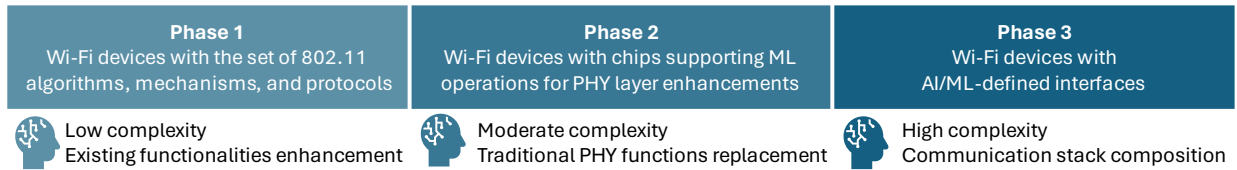

Figure 3. AI/ML-defined radio roadmap

## 3.2. Standardization Efforts

Standardization bodies will play a crucial role in the adoption of AI/ML-defined radios. Similarly as nowadays, where the implementation of the various components is left open, the algorithms driving the operational logic, the responsibility for developing ML models to replace traditional algorithms and specific solutions will largely fall to chip and equipment manufacturers. Nevertheless, to fully leverage AI/ML capabilities, standardization bodies will need to define new elements to support the implementation of AI/ML solutions, including the establishment of new interfaces and protocols for data collection, ML model training, and ML model exchange. Additionally, standardized testing and validation processes will be required to ensure not only interoperability among ML-capable devices but also backward compatibility and fairness with non-AI/ML-defined radios when new ML functions are introduced. Lastly, and perhaps the most important aspect, standardization bodies must implement mechanisms that provide trust on ML model decisions so no devices can misbehave.

In recent years, several standardization bodies have initiated efforts towards advancing AI/ML-defined radios by releasing studies and documentation. Key contributions have come from ITU-T, including ITU-T Y.3172 on flexible architectures for ML-aware networks [22] and ITU-T Y.3181 on an architectural framework for ML Sandbox [20]. Additionally, 3GPP has contributed with 3GPP Rel. 18 focusing on the study of an AI-native air interface, and 3GPP Rel. 19, expected by 2026, which already targets to implement an AI-native air interface. The IEEE 802.11 community has also contributed through its AIML TIG technical report [14]. Specifically, for Wi-Fi, the IEEE 802.11 TIG has highlighted several representative use-cases where ML models can offer superior performance compared to traditional approaches. These include CSI feedback compression using NNs, enhanced roaming assisted by AI/ML, deep reinforcement learning (DRL) for improved channel access, and MAPC driven by AI/ML.

Will Wi-Fi 8 be the first 802.11 amendment supporting AI/ML operations? We do not expect that the IEEE 802.11bn Task Group will standardize any particular aspect related to AI/ML since a quasi-definitive agreement of the potential features to be included in Wi-Fi 8 will be made early next year (the Spec Framework Document, SFD, is expected for January 2025 and Draft 1.0 for May 2025), and AI/ML topics have not yet been studied enough to be elected as a candidate feature for the standard. However, it is likely that chip and equipment vendors manufacturers will start including ML models in their proprietary radio firmwares. However, based on the recent establishment of the AIML Standing Committee within the IEEE 802.11 Working Group, an important boost in AI/ML standardization for Wi-Fi can be expected for the following years.

## 4. Conclusion

In recent years, researchers have been applying ML-based methods to improve the performance of existing wireless networks. The next step in the evolution of wireless radio interfaces is to have native support for AI/ML-based decision-making within communication functions. Based on the standardization activities, industry trends, and our own research activities reported above, we foresee that such AI/ML-defined radios will be part of the next generation of wireless communications systems.

## Acknowledgements

This paper is supported by the CHIST-ERA Wireless AI 2022 call MLDR project (ANR-23-CHR4-0005), partially funded by AEI and NCN under projects PCI2023-145958-2 and DEC-2023/05/Y/ST7/00004, respectively. In addition, B. Bellalta's contribution is also supported by Wi-XR PID2021123995NB-I00 (MCIU/AEI/FEDER,UE) and MdMCEX2021-001195-M/AEI /10.13039/501100011033. For the purpose of Open Access, the author has applied a CC-BY public copyright licence to any Author Accepted Manuscript (AAM) version arising from this submission.

## Biographies

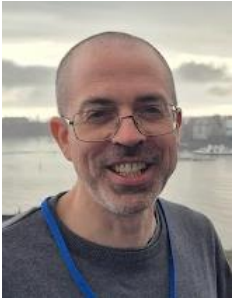
**Boris Bellalta** (SM'13) is a full professor at Universitat Pompeu Fabra (UPF), where he heads the Wireless Networking group. His research interests are in the area of wireless networks and performance evaluation, with emphasis on Wi-Fi technologies, and Machine Learning-based adaptive systems.

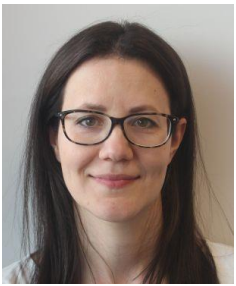
**Katarzyna Kosek-Szott** is a full professor at AGH University of Krakow, Poland. Her research interests are in the area of wireless networks, with emphasis on novel 802.11 mechanisms, performance improvement with machine learning, and the coexistence of wireless technologies in unlicensed bands.

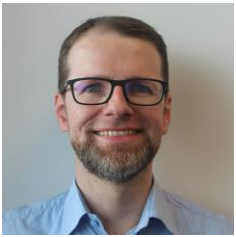
**Szymon Szott** is an associate professor at AGH University of Krakow, Poland. His professional interests are related to wireless local area networks (channel access, quality of service, security, inter-technology coexistence). He is a voting member of the IEEE 802.11 Working Group.

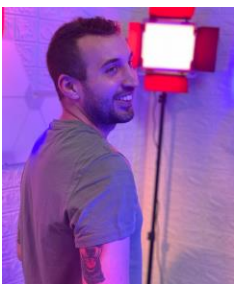
**Francesc Wilhelmi** (M'23) is a researcher at Nokia Bell Labs. His main research interests are Wi-Fi technologies and their evolution, network simulators and network digital twinning, machine learning, decentralized learning, and distributed ledger technologies.